\documentclass[letterpaper]{article} % DO NOT CHANGE THIS
\usepackage[submission]{aaai24}  % DO NOT CHANGE THIS
\usepackage{times}  % DO NOT CHANGE THIS
\usepackage{helvet}  % DO NOT CHANGE THIS
\usepackage{courier}  % DO NOT CHANGE THIS
\usepackage[hyphens]{url}  % DO NOT CHANGE THIS
\usepackage{graphicx} % DO NOT CHANGE THIS
\usepackage{natbib}  % DO NOT CHANGE THIS AND DO NOT ADD ANY OPTIONS TO IT
\usepackage{caption} % DO NOT CHANGE THIS AND DO NOT ADD ANY OPTIONS TO IT
\usepackage{amsmath}
\usepackage{amssymb}
\usepackage{makecell}
\usepackage{algorithm}
\usepackage{algorithmic}

\usepackage{newfloat}
\usepackage{listings}
\DeclareCaptionStyle{ruled}{labelfont=normalfont,labelsep=colon,strut=off} % DO NOT CHANGE THIS
\floatstyle{ruled}
\newfloat{listing}{tb}{lst}{}
\floatname{listing}{Listing}
\title{Learning to Control Autonomous Fleets from Observation \\ via Offline Reinforcement Learning}
\author {
    Carolin Schmidt\textsuperscript{\rm 1},
    Daniele Gammelli\textsuperscript{\rm 2},
    Francisco Camara Pereira\textsuperscript{\rm 1}
    Filipe Rodrigues\textsuperscript{\rm 1}
}
\affiliations {
    \textsuperscript{\rm 1}Technical University of Denmark, Kongens Lyngby, Denmark\\
    \textsuperscript{\rm 2}Stanford University, Stanford, California\\
    csasc@dtu.dk, gammelli@stanford.edu, camara@dtu.dk, rodr@dtu.dk
}

\usepackage{bibentry}
\begin{document}

\maketitle

\begin{abstract}
Autonomous Mobility-on-Demand (AMoD) systems are an evolving mode of transportation in which a centrally coordinated fleet of self-driving vehicles dynamically serves travel requests.
The control of these systems is typically formulated as a large network optimization problem, and reinforcement learning (RL) has recently emerged as a promising approach to solve the open challenges in this space.
Recent centralized RL approaches focus on learning from online data, ignoring the per-sample-cost of interactions within real-world transportation systems.
To address these limitations, we propose to formalize the control of AMoD systems through the lens of offline reinforcement learning and learn effective control strategies using solely offline data, which is readily available to current mobility operators.
We further investigate design decisions and provide empirical evidence based on data from real-world mobility systems showing how offline learning allows to recover AMoD control policies that (i) exhibit performance on par with online methods, (ii) allow for sample-efficient online fine-tuning and (iii) eliminate the need for complex simulation environments.  
Crucially, this paper demonstrates that offline RL is a promising paradigm for the application of RL-based solutions within economically-critical systems, such as mobility systems.
\end{abstract}

\section{Introduction}
As the world's population continues to urbanize, with projections indicating that over $60\%$ of the population will reside in urban environments by 2050, there is an urgent need for innovative mobility solutions \cite{UNDepEconAff2021}. The traditional model of urban transportation, heavily reliant on private cars, is no longer sustainable and is widely recognized as the main cause of increasingly congested transportation systems. Within this context, Autonomous Mobility-on-Demand (AMoD) systems have the potential to transform urban transportation by providing cheap and efficient point-to-point trips while reducing the need for private car ownership. In an AMoD system, the customer requests a one-way ride from their origin to a destination and is matched with an autonomous vehicle belonging to a larger fleet. In real-world systems, the effectiveness of rebalancing strategies is central to the overall system performance, with sub-optimal strategies potentially exacerbating congestion through unnecessary trips or increased passenger waiting times. In practice, AMoD systems enable the centralized control of the fleet, thus potentially allowing for extremely efficient transportation systems. However, the task of managing and routing a large number of vehicles within real-world transportation systems is typically formulated as a large network optimization problem, which may be prohibitively expensive in practice, and its solution is considered an open problem.

Recently, reinforcement learning (RL) has emerged as a promising approach to solving the AMoD control problem while avoiding expensive optimization routines \cite{MAO2020, Feng, Gammelli1, Gammelli2}. A traditional approach to RL typically entails the definition of an agent that gradually improves its performance by repeatedly interacting with an environment. However, the necessity to learn from online data is also one of the major constraints to their real-world adoption, especially within safety or economically-critical systems \cite{Levine2020} (e.g., healthcare,  autonomous driving, etc.). An alternative approach is to train via simulation, which can be summarized as the process of (i) building a complex simulator of urban mobility, (ii) training AMoD controllers in simulation, and (iii) deploying the agent in the real world \cite{Candela}. However, these approaches require access to reliable simulators -- in itself a challenging task -- and are potentially exposed to unexpected sim-to-real distribution shifts \cite{8793789}. In this work, we argue that offline RL represents a promising direction to overcome these challenges by learning from real, pre-collected data. Crucially, online interaction within real-world AMoD systems is either extremely expensive or infeasible, while historical data from service operators is likely to be abundant and readily available. Thus, we propose offline RL as a framework to enable service operators to completely avoid expensive fleet management decisions until a sufficient level of performance is guaranteed, allowing for further online fine-tuning once the policy is safer and cheaper to deploy and taking a step toward the deployment of learning-based solutions for the system-level control of real-world transportation systems.\\

The contributions of this paper are threefold:
\begin{itemize}
 \item We formulate the AMoD control problem through the lens of offline RL and propose an approach that enables RL agents to centrally control AMoD systems from solely offline data. 
  \item We investigate design decisions within our framework, such as the relation between dataset quality and coverage, and quantify their impact on model performance and transferability. 
  %\item We show that our approach achieves performance on par with methods based on online learning while drastically improving data efficiency, ultimately providing a practically-feasible strategy to deploy RL algorithms within real-world mobility systems.
  \item We show that our approach learns policy initializations that enable sample-efficient online fine-tuning, ultimately providing a practically-feasible strategy to deploy RL algorithms within real-world mobility systems. 
\end{itemize}

\section{Related Work}
Existing literature on the AMoD control problem can be broadly classified into rule-based heuristics \cite{Hyland, LEVIN2017373}, model predictive control (MPC) methods \cite{Zhang, Iglesias}, and RL approaches. Readers interested in a recent survey about centralized and decentralized RL approaches for vehicle relocation can refer to \citet{RLSurvey}. 

As AMoD systems enable the centralized control of the fleet, we focus on reviewing system-level vehicle rebalancing with central decision-making. Such methods aim to explicitly modify the current distribution of idle vehicles to collectively improve system performance and serve more requests. Due to the intractability of the joint action space for all vehicles, centralized RL approaches for the control of ride-hailing systems have only been studied recently. To tackle this challenge of scalability, \citet{MAO2020} devise a system policy that specifies the aggregated number of vehicles to be rebalanced between zones, \citet{Feng} formulate a policy that determines a sequence of atomic actions, and \citet{Fluri} deploy a cascaded multi-level Q-learning approach. \citet{DENG2022} train a policy that specifies a probability distribution for rebalancing idle vehicles to their k-nearest neighbors. \citet{Gammelli1} propose a graph neural network-based Dirichlet policy that determines the desired idle vehicle distribution in the network. Similarly, \citet{Filipovska} deploy a policy that specifies a Dirichlet distribution of vehicles to be rebalanced between zones. 

Overall, approaches based on RL demonstrate computational tractability without significantly sacrificing optimality. However, despite their effectiveness, the aforementioned studies do not consider the cost of training a reinforcement learning agent from scratch, which can be extremely high within real-world systems (i.e., consider the monetary cost of trying different fleet rebalancing strategies for the purpose of RL exploration). No publicly known simulation environment for ride-hailing services has demonstrated sufficient fidelity to the real world for direct sim-to-real transfer \cite{RLSurvey}. Thus, deployed work has either adopted offline \cite{Jiao201, Tang2021} or fully-online learning of value functions \cite{Han2022, Eshkevari2022}. %\citet{Jiao201} combine offline learning and online planning i.e. decision-time planning, and \citet{Tang2021} propose an ensemble approach where the offline trained value function is updated during the online phase. \citet{Han2022, Eshkevari2022} argue for a fully-online approach, wherein the value function is learned on-policy from streaming events. 
To address the challenges of the high-dimensionality state space, these studies assume that the global value function can be decomposed into the individual drivers' value functions. Such an assumption is valid where individual goals align with the global system's objective, e.g., in vehicle dispatching. 
%This assumption does not impact optimality if global maximization can be achieved through individual maximization, as is the case with vehicle dispatching/matching. 
However, for vehicle rebalancing, this assumption does not hold and would lead to idle drivers clustering at a single high-value location. To mitigate this undesirable behaviour, \citet{Jiao201} and \citet{Tang2021} sample a destination for each driver from a Boltzmann distribution. \citet{Han2022, Eshkevari2022} only consider vehicle dispatching, but their approach could also be extended accordingly. This method avoids the need for an explicit policy but may compromise system performance due to limited coordination. From the AMoD service provider's perspective, the elimination of drivers renders individual objectives obsolete, placing the focus on system-wide objectives for successful operations. %From the AMoD service provider's perspective, the absence of drivers leaves their objectives obsolete and system-wide objectives become the center of successful operations. %For an AMoD service provider, explicit consideration of system-wide objectives becomes paramount, as individual driver objectives are not existent anymore. %Limited coordination alludes to a scenario where individual vehicles operate based on individual objectives without explicit consideration of system-wide objectives. 
A solution lies in a centralized parametric policy, which aligns vehicle actions with system goals by explicitly optimizing for system performance as in \cite{MAO2020, Feng, Fluri, Gammelli1, DENG2022, Filipovska}. However, if aiming at deploying a parametric policy, fully-online approaches \cite{Han2022, Eshkevari2022} could not be deployed, as an untrained policy could incur significant real-world costs when interacting with the AMoD system. In light of this, learning a policy from an offline dataset requires explicit offline RL methods \cite{Levine2020}, while previous offline approaches \cite{Jiao201, Tang2021} focuse solely on deriving value estimates from the dataset. In this work, we aim to address these challenges by proposing offline RL to learn a policy for system-level rebalancing as a mathematical framework to eliminate the need for online learning from scratch and learning in simulation environments.
%An explicit parametric policy provides a remedy through centralized, system-level decision-making. This aligns the actions of the vehicle with the overall system's objectives. If providers seek explicit coordination through centralized decision-making, they must optimize for it, i.e. employ a parametric policy, as in \cite{MAO2020, Feng, Fluri, Gammelli1, DENG2022, Filipovska}. This renders a fully online approach untenable, as letting an untrained policy interact with the AMOD system will almost certainly result in high costs for sub-optimal decisions. Therefore, it is essential to ensure that the policy is pre-trained before deployment in a real-world scenario. In this work, we aim to address these challenges by proposing offline RL to learn a policy for system-level rebalancing as a mathematical framework to eliminate the need for online learning from scratch and learning in simulation environments.

Applications of offline RL can be found in diverse research fields, such as robotics \cite{kumar2021a}, healthcare \cite{Wang2018}, recommender systems \cite{Gilotte}, and more. Within the transportation domain, most literature focuses on autonomous driving \cite{Guan2021, Shi2021, Fang2022}, road traffic control \cite{Kunjir2022a, Dai2021} and pricing for ride-hailing services \cite{wu2022spatiotemporal}. 
%Related work for mobility-on-demand systems \cite{Jiao201} proposes a value-based approach for a ride-hailing platform with drivers, where a value is learned for each cell in a hexagon grid system. They learn the value function offline from pre-collected data with tabular reinforcement learning and SARSA, estimating the value of the behavior policy used to collect the dataset. Rebalancing actions are then chosen according to the values of the neighboring cells. To avoid crowding all idle drivers into a single location with the highest value, they add stochasticity to the action by sampling from a Boltzmann distribution. To address the limitation that this offline learned value function cannot adapt to changing marketplace conditions, \cite{Tang2021} proposes a periodic ensemble method, combining online learning with offline priors. Finally, \cite{Han2022} proposes a fully online approach that learns the value function from streaming events while keep being updated during deployment to adapt to changing marketplace conditions. Although \cite{Han2022} only consider the dispatching problem, their approach could include the vehicle rebalancing problem in a similar fashion as \cite{Jiao201} and \cite{Tang2021}. 
However, to the best of our knowledge, no research study used offline RL for the system-level control of large fleets of vehicles with central decision-making.
In this work, we develop an offline RL framework that leverages the specific strengths of direct optimization and reinforcement learning to control fleets of vehicles within urban transportation networks.
\section{Background}
In this section, we introduce the notation and theoretical background underlying our work in the context of RL, offline RL, conservative Q-learning, and online fine-tuning. 

\subsection{The Reinforcement Learning Problem}\label{subsec:RL}
In reinforcement learning, we aim to learn to control a dynamic system from experience. Specifically, we consider a (fully-observable) infinite-horizon Markov decision process (MDP) $M = (S, A, P, d_{0}, r, \gamma)$, where $S$ is a set of possible states $s \in S$, $A$ is a set of possible actions $a \in A$, $P$ defines the conditional probability distribution $P(s_{t+1}|s_{t},a_{t})$ that specifies the dynamics of the system, $d_0$ defines the initial state distribution $d_0(s_0)$, $r:S \times A \rightarrow \mathbb{R} $ describes a reward function and $\gamma \in (0,1]$ is a scalar discount factor.
The goal of an RL agent is to learn a policy, which defines a distribution over possible actions conditioned on the state $\pi(a_{t}|s_{t})$ by interacting with the MDP $M$ (i.e., the environment) under the objective of maximizing the expected sum of cumulative rewards. %Hence, the general RL objective can be derived as $J(\pi) = \mathbb{E}_{(s_t, a_t) \sim \pi } [\sum_{t=0}^{T} \gamma^tr(s_{t}, a_{t})]$ for time horizon $\mathcal{I}=\{1,2,...,T\}$. 
%This objective can be extended to a more general maximum entropy objective \cite{Ziebart} with
%\small 
%\begin{equation}
%\pi^*=\arg \max_{\pi} \mathbb{E}_{(s_t,a_t) \sim \pi}\Bigl[\sum_{t=0}^{T} \gamma^t\Bigl(r\left(s_t, a_t\right)+\alpha \mathcal{H} \left(\pi\left(\cdot \mid s_t\right)\right)\Bigr)\Bigr], \nonumber
%\end{equation}
%\normalsize
%which has been shown to improve exploration and speed of %convergence \cite{Haarnojab}.
An approach of particular interest for this work is the Soft Actor-Critic (SAC) \cite{Haarnojab} algorithm, which combines an actor-critic formulation with maximum entropy reinforcement learning. We consider a parametric policy $\pi_\phi(a_t|s_t)$, i.e., the ``actor'', that maps states to actions, and a Q-function $Q_{\theta}(s_t, a_t)$, i.e., the ``critic'', that estimates the expected return when starting from state $s$, taking an arbitrary action $a$, and then continuing acting according to policy $\pi$.
% \begin{equation}
%\begin{aligned}
%Q^\pi(s, a)=&\underset{(s_t,a_t)\sim \pi}{\mathbb{E}}[\sum_{t=0}^{\infty} \gamma^t r(s_t, a_t)+ \\ &\alpha %\sum_{t=1}^{\infty} \gamma^t H(\pi(\cdot \mid s_t)) \mid s_0=s, a_0=a].\nonumber
%\end{aligned}
%\end{equation}
%The SAC algorithm is an off-policy algorithm, such that the agent continuously interacts with the environment but updates its policy based on a collection of past transitions $(s_t, a_t, s_{t+1}, r_{t})$, often referred to as a replay buffer $\mathcal{D}$. Off-policy methods have been shown to be more sample-efficient as data is not discarded at the end of an episode, as in on-policy methods. Specifically, SAC is based on soft policy iteration, thus alternating between policy evaluation and policy improvement steps. In the policy evaluation step, 
The Q-value is optimized to approximate the expected return under the current policy $\pi$. Specifically, the Q-values are updated by iteratively applying the soft Bellman backup operator $B^{\pi}$ given by:
\begin{equation}
\begin{aligned}
B^{\pi} Q(s_t,a_t) &= r(s_t, a_t)+ \gamma \mathbb{E}_{s_{t+1} \sim P} [\mathbb{E}_{a_{t+1} \sim \pi} \\ & [ Q(s_{t+1}, a_{t+1})- \alpha \log (\pi_\phi(a_{t+1} \mid s_{t+1}))]], \nonumber
\end{aligned}
\end{equation}
where $(s_t, a_t)$ is sampled from a collection of past transitions $\mathcal{D} =\{(s_t, a_t, s_{t+1}, r_{t})\}$, often referred to as a replay buffer. %Thus, we optimize the Q-function to minimize the Bellman residual by:
%\begin{equation}
%\begin{aligned}
%&J_Q(\theta) = \mathbb{E}_{(s_t, a_t, s_{t+1})\sim  \mathcal{D}} \Bigl[\frac{1}{2} \Bigl(Q_\theta(s_t, a_t) - \\&(r(s_t, a_t) + \gamma  \mathbb{E}_{a_{t+1} \sim \pi} [Q_{\theta}(s_{t+1}, a_{t+1}) - \alpha \log (\pi_\phi(a_{t+1} \mid s_{t+1})) ])\Bigr)^2\Bigr], \nonumber
%\hat{\nabla}_\theta J_Q(\theta) & =\nabla_\theta Q_\theta(s_t, a_t)\Bigl(Q_\theta(s_t, a_t)- \\ & \Bigl(r(s_t, a_t)+\gamma(Q_{\bar{\theta}}(s_{t+1}, a_{t+1})-\alpha \log (\pi_\phi(a_{t+1} \mid s_{t+1})))\Bigr)\Bigr).\nonumber
%\end{aligned}
%\end{equation}
The policy is updated to maximize the expected return, represented by the Q-values, alongside the entropy of the policy: 
\begin{equation}
J_\pi(\phi)=\mathbb{E}_{s_t \sim \mathcal{D}}\left[\mathbb{E}_{a_t \sim \pi_\phi}\left[\alpha \log \left(\pi_\phi\left(a_t \mid s_t\right)\right)-Q_\theta\left(s_t, a_t\right)\right]\right].\nonumber
\end{equation}

\subsection{Offline Reinforcement Learning}
\label{subsec:OfflineRL}
Offline RL is a learning paradigm that involves learning on a static dataset of transitions $\mathcal{D}=\{(s_t^{i},a_t^{i},s_{t+1}^{i},r_t^{i})\}$ collected by a potentially sub-optimal policy $\pi_{\beta}$.
%Theoretically, any off-policy algorithm could be applied to learn a policy from offline data. However, these methods often fail because they were developed under the assumption that erroneous estimates (e.g., estimates of Q-values) can be corrected by online data collection. Without the active exploration necessary to correct this erroneous behavior, conventional off-policy RL approaches are often unable to learn an effective strategy from offline data. 
The fundamental challenge of offline RL is to find a strategy that is different from the behavior policy $\pi_{\beta}$ and, at the same time, avoids erroneous behavior outside the data distribution. For a comprehensive review of offline RL approaches, interested readers can refer to \cite{Levine2020, Prudencio2022}. %At a high level, offline RL approaches can be categorized into model-based \cite{MOPO, MOREL}, model-free \cite{Kumar,AWR,BCQ,BEAR}, and recently, trajectory optimization approaches \cite{DT, TT}, depending on whether data is used to learn a model of the dynamics, to learn a policy or to learn a trajectory distribution \cite{Prudencio2022}. For a comprehensive review, interested readers can refer to \cite{Levine2020, Prudencio2022}.
%For example, model-based approaches learn to estimate a model of the environment and the reward function, which are later used to improve the policy. These models typically make use of uncertainty estimates to penalize the reward function in out-of-distribution states. In model-free offline RL, policy constraint methods directly or implicitly constrain the learned policy $\pi$ to be close to the behavior policy $\pi_{\beta}$ under some distance metric (e.g., KL-divergence). 
%Value regularization methods attempt to learn a lower bound of the true value function to achieve a conservative estimate of the expected reward. For a comprehensive review, interested readers can refer to \cite{Levine2020, Prudencio2022}.

In this work, we exploit ideas from the conservative Q-learning algorithm (CQL) \cite{Kumar}. In Q-learning, querying the value function on out-of-distribution actions that are not forced to obey the Bellman residual typically leads to an overestimation of Q-values. As introduced in the CQL algorithm, this overestimation can be solved via regularization of the Q-function, concretely incentivizing a conservative estimate of the true Q-function by minimizing Q-values in addition to the standard Bellman error.
%The CQL objective is composed of three parts: 1) the standard Bellman residual, which ensures that Q-values for in-distribution actions are accurate, 2) a regularization term that penalizes high Q-values for unseen actions at a given state, and 3) an additional maximization term under the data distribution which prevents excessive underestimation of in-distribution action values.
Formally, the training objective of CQL, in addition to the standard bellman residual, is given by: 
\begin{equation}
\label{eqn:CQL}
\min_{Q} \eta \,\mathbb{E}_{s\sim \mathcal{D}, a\sim \pi} [Q(s,a)] - \mathbb{E}_{s, a\sim \mathcal{D}}[Q(s,a)]\nonumber,
\end{equation}
where $\eta$ is the regularizer weight that controls the trade-off between the standard Bellman residual and the CQL regularization term. With a suitable $\eta$, this formulation has proven to be an effective and competitive way of training an agent offline \cite{Kumar, D4RL, qin2022neorl}. 

With offline RL, we obtain a policy initialization, which is intended for sample-efficient online fine-tuning. However, conservative methods tend to learn smaller Q-values than their true values. Consequently, initial interactions during online fine-tuning are spent adjusting the Q-function, leading to unintentional unlearning of the initial policy. To address this, \citet{nakamoto2023calql} propose calibrating the Q-function during offline learning to match the range of the ground-truth Q-values. Their approach Cal-CQL ensures the learned, conservative Q-values are lower-bounded by the ground-truth Q-values of a sub-optimal reference policy, denoted as $\mu$, for which we can choose the behavior policy of the dataset. %Conveniently, the behavior policy of the dataset can be chosen as this reference policy. 
Practically, this is achieved by masking out the push-down on the Q-value for out-of-distribution (OOD) actions if they are not calibrated. This modifies the CQL regularizer to:
\begin{equation}
\min_{Q} \eta \, \mathbb{E}_{s\sim \mathcal{D}, a\sim \pi} [max(Q(s,a), V^\mu(s))] - \mathbb{E}_{s,a\sim \mathcal{D}} [Q(s,a)] \nonumber.
\end{equation}

\section{The AMoD Problem}
We now define the terminology associated with the AMoD control problem. An on-demand service provider coordinates \textit{M} single-occupancy autonomous vehicles on a transportation network represented by a complete graph $\mathcal{G}=(\mathcal{V}, \mathcal{E})$ where $\mathcal{V} = \{v_i\}_{\{i=1:N_v\}}$ and $\mathcal{E}=\{e_j\}_{\{j=1:Ne\}}$ represent the set of vertices and edges of $\mathcal{G}$. Specifically, $\mathcal{V}$ defines the set of stations (e.g., pick-up or drop-off locations), and $\mathcal{E}$ defines the shortest paths between stations. The time horizon is discretized into a set of time steps $\mathcal{I} = {1,2,...,T}$ of length $T$. At any time step $t$, vehicles are controlled to travel along the shortest path between station $i$ and $j\neq i \in \mathcal{V}$ with a travel time of $\tau^t_{i,j} \in \mathbb{Z}^+$ and travel cost $c_{ij}$, as a function of travel time. At each time step $t$, passengers submit trip requests for a desired origin-destination pair $(i,j) \in \mathcal{V} \times \mathcal{V}$, which is characterized by demand $d^t_{i,j}$ and price $p^t_{i,j}$. The operator matches passengers to vehicles, and the vehicles will transport the passengers to their destinations. For idle vehicles that are not matched with any passengers, the operator controls them to stay at the same station or rebalance to other stations. We denote $x^t_{i,j} \in \mathbb{N}, x^t_{i,j} \leq  d^{t}_{i,j}$ as the passenger flow, i.e., the number of passengers traveling from station $i$ to station $j$ at time step $t$ and $y^t_{i,j} \in \mathbb{N}$ as the rebalancing flow, i.e., the number of vehicles rebalancing from station $i$ to station $j$ at time step $t$.

\section{Offline Reinforcement Learning for AMoD}
In this section, we formulate the AMoD rebalancing MDP and formalize Conservative Q-learning for AMoD systems.
\begin{figure*}[t]
    \centering
    \includegraphics{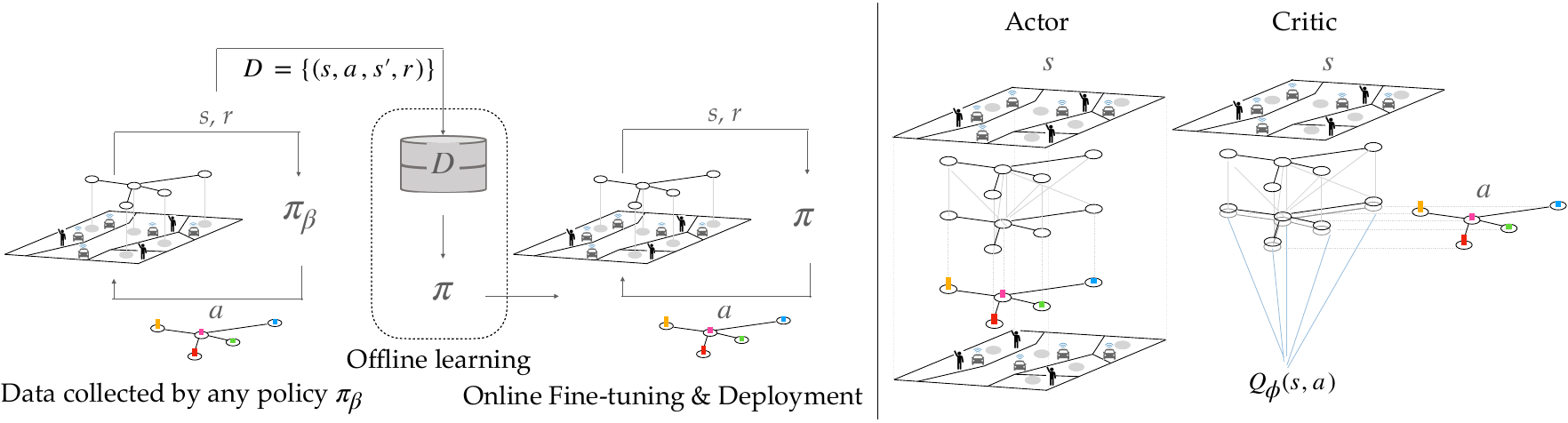}
    \caption{(Left) The policy $\pi$ is trained on a dataset collected by a (potentially sub-optimal) policy $\pi_{\beta}$ without any environment interaction before being deployed. (Right) Both actor and critic update raw graph representations of the transportation network to compute (i) a desired distribution of idle vehicles \textit{a} and (ii) an estimate of the Q-function, respectively.}
    \label{fig:offlinerl}
\end{figure*}

 \subsection{The AMoD Rebalancing MDP}
 \label{subsec:MDP}
 As in \cite{Gammelli1, Gammelli2, gammelli2023graph}, this paper adopts a three-step decision-making framework to tackle the AMoD control problem. This framework comprises three stages: (1) solving a matching problem to derive the passenger flow, (2) determining the desired distribution of idle vehicles through the use of the learned policy $\pi_\phi(a_t|s_t)$, (3) converting this distribution to a rebalancing flow by solving a minimal rebalancing-cost problem. This approach has been shown to outperform pure end-to-end RL approaches \cite{gammelli2023graph}, and we provide more details about steps 1) and 3) in the Appendix. The advantage of this procedure is the reduction of the action space from $|\mathcal{V}|^2$ to $|\mathcal{V}|$, as the policy defines the action at each station as opposed to along each edge (i.e., each origin-destination pair).
Our goal is to learn a policy to compute the desired distribution of idle vehicles (Step 2) via offline data. However, we want to emphasize that our approach is agnostic to the framework and can be used as an indication for other approaches that learn  rebalancing policies. We define the AMoD rebalancing problem as an MDP $M^{reb} = (S^{reb}, A^{reb}, P^{reb}, d_0^{reb}, r^{reb}, \gamma)$ characterized by the following elements:

\smallskip\noindent\textbf{Reward ($r^{reb}$)}: we choose the reward to be the operator's profit, which we define as the difference between the revenue from serving passengers and the cost of operations:
\begin{equation}
r^t_{reb} = \sum_{i,j \in \mathcal{V}} x_{ij}^{t+1}(p^{t+1}_{ij}-c^{t+1}_{ij}) -  \sum_{(i,j) \in \mathcal{E}} y_{ij}^t c_{ij}^t. \nonumber
\label{eq:reward}
\end{equation}
%An important insight in Eq \eqref{eq:reward} is that the reward at time $t$ is defined as a function of values at time $t+1$ for all passenger-related elements (i.e. $x_{ij}^{t+1}, \, p^{t+1}_{ij}, \, c^{t+1}_{ij}$), and at time $t$ for all rebalancing-related elements (i.e., $y_{ij}^t, \, c_{ij}^t$).
%Within our problem formulation, the agent receives the current state $s_t$ after the assignment of passengers at time $t$ and generates an action $a_t$, which then leads to rebalancing trips, also at time $t$. At the next time step, passengers are assigned to idle vehicles, thus resulting in $s_{t+1}$.

\smallskip\noindent\textbf{Action space ($A^{reb}$)}: given the number of idle vehicles and their current spatial distribution, we consider the problem of determining the \textit{desired idle vehicle distribution} $a^t_{reb}$. Specifically, the policy describes a probability distribution over stations, indicating the percentage of idle vehicles to be rebalanced towards each station.
 
\smallskip\noindent\textbf{State space ($S^{reb}$)}: we define the state to contain the information needed to determine proactive rebalancing strategies, including the structure of the transportation network through its adjacency matrix \textbf{A} and state-level information by means of a feature matrix \textbf{X}.
Specifically, given a planning horizon $K$, we consider: (1) the current and projected availability of idle vehicles in each station $m^t_{i} \in [0,M], \forall i \in \mathcal{V}$ and ${\{m^{t'}_{i,j}\}}_{t'=t,...,t+K}$, (2) provider-level information trip price $p^t_{i,j}$ and cost $c^t_{i,j}$, (3) current $d^t_{ij}$ and estimated ${\{{\hat{d}^{t'}_{i,j}}\}}_{t'=t,...,t+K}$ transportation demand between all stations. We assume to have access to a noisy and unbiased estimate of demand in the form of the rate of the underlying time-dependent Poisson process describing travel behavior in the system, although this could come from a prediction model such as \cite{Geng_Li_Wang_Zhang_Yang_Ye_Liu_2019}

\smallskip\noindent\textbf{Dynamics ($P^{reb}$)}: The stochastic evolution of travel demand between stations follows a time-dependent Poisson process, with a time-varying arrival rate estimated from real data. The process is independent of the rebalancing action and the arrival process of other passengers in other locations. Additionally, the evolution of state elements is characterized as follows: (1) the estimated availability ${\{{m^{t'}_{i,j}}\}}_{t'=t,...,t+K}$ is determined based on the number of idle vehicles at station $i$ at time step $t+1$ plus the number of incoming vehicles minus the number of outgoing vehicles (from both passenger and rebalancing flow), (2) provider-level state variables, such as trip price $p^t_{ij}$ and cost $c^t_{ij}$, are assumed to be exogenous and known beforehand.\\

\begin{table*}
  \begin{tabular}{l|ccc|cc|cc}
      \hline
    City&Random&No Reb.&ED&CQL&SAC&MPC-Forecast&MPC-Oracle\\
      \hline
    NYC Brooklyn &26.5($\pm{0.4}$)&27.6($\pm{0.6}$)&49.2($\pm{0.9}$)&53.1($\pm{1.1}$)&56.1($\pm{1.1}$)&\textbf{56.5}($\pm{1.2}$)&57.2($\pm{0.8}$)\\
    Shenzhen West &48.6($\pm{1.7}$)&60.0($\pm{0.7}$)&58.4($\pm{0.9}$)&61.6$(\pm{0.5})$&\textbf{62.3}($\pm{0.7}$)&60.6($\pm{1.0}$)&65.2($\pm{1.0}$)\\
    San Francisco &12.5($\pm{0.6}$)&10.1($\pm{0.4}$)&14.1$(\pm{0.4})$&14.2$(\pm{0.4})$&\textbf{14.8}($\pm{0.3}$)&13.9($\pm{0.5}$)&15.9($\pm{0.4}$)\\
    \hline
\end{tabular}
  \caption{Average reward (profit, thousands of dollars). Bold highlights best performing (non-oracle) model.}
    \label{tab:tab1}
\end{table*}

\subsection{Conservative Q-Learning for AMoD Systems}
 \label{subsec:CQL_for_amod}
We propose a SAC formulation with the conservative loss of CQL as an approach to learning rebalancing policies offline. We assume the availability of a dataset $\mathcal{D}=(s_t, a_t, s_{t+1},r_t)$ containing historical rebalancing decisions collected by a behavior policy $\pi_{\beta}$. Our objective is to train an agent using this dataset without any further interactions with the AMoD system and to achieve a policy that matches or even outperforms $\pi_{\beta}$. The offline learning process and the actor and critic architectures are illustrated in Figure \ref{fig:offlinerl}. We use GNN encoders to parameterize the neural architecture for the policy $\pi_\phi\left(a_{t} | s_{t}\right)$ and the Q-function $Q_{\theta}(s_t, a_t)$. GNNs have proven to be particularly effective in transportation applications due to their ability to capture and process spatial relations within the network. Additionally, GNNs can process transportation networks of varying sizes and connectivity, while traditional machine learning methods, such as MLPs, are limited to fixed-size inputs. In what follows, we introduce the neural network architectures in more detail:

\smallskip\noindent\textbf{Policy network $\pi_\phi\left(a | s\right)$}. To define a valid vehicle distribution, the output of the policy network is sampled from a Dirichlet distribution. More precisely, the network architecture comprises one graph convolutional layer with skip-connections that uses a sum-pooling function as neighborhood aggregation and a ReLU non-linearity on its output. This is followed by three MLP layers (with ReLU activations on the first two) that output the concentration parameter $\textit{c}$ for the Dirichlet distribution. To ensure the positivity of $\textit{c}$, we apply a softplus activation function in the last layer. 

\smallskip\noindent\textbf{Critic network $Q_{\theta}(s_t, a_t)$}. We propose an architecture for the critic which uses the same GNN encoder as the policy network. The main difference to the actor architecture is that the encoded state information is concatenated with the action on a node level, followed by two MLP layers with ReLU activation. The global sum-pooling function is not performed until before the last MLP layer. We found that this architecture choice is essential for obtaining satisfactory results.  

\section{Experiments}
In this section, we present simulation results using data from different cities. Specifically, the goal of our experiments is to answer the following questions: (1) can a policy solely trained on a static dataset learn effective rebalancing strategies in real-world urban mobility scenarios? (2) how do characteristics of the dataset influence the performance and transfer capabilities of offline RL agents? and (3) can we quantify the benefits of deploying an agent trained offline compared to one trained from scratch via online data?(Code available: \emph{\url{https://github.com/carolinssc/offline-rl-amod}})

\subsection{Benchmarks}
In our experiments, we compare our proposed offline-RL framework with the following methods:

\smallskip\noindent\textbf{No rebalancing}: we measure the performance of the system without any rebalancing activities. %This baseline can be regarded as a lower bound for the system, as any effective rebalancing activity should lead to an overall improvement.

\smallskip\noindent\textbf{Heuristics}: 
\begin{enumerate}
\item \textit{Random policy}: at each time step, the desired distribution is sampled from a Dirichlet prior with a concentration parameter $\textit{c} = [1,1,...,1]$.
\item \textit{Equally distributed policy} (ED): rebalancing actions are selected to restore an equal distribution of vehicles across all areas in the transportation network.
\end{enumerate}
\smallskip\noindent\textbf{Learning-based}: within this class of methods, we measure the performance of RL agents trained via online learning.
\begin{enumerate}
\item \textit{SAC}: the online version of our method based on the Soft Actor-Critic algorithm. In practice, this benchmark serves as an upper bound of performance for our offline approach as we define both actor and critic networks as in our CQL formulation, but with the additional possibility of infinitely interacting with the environment.
\end{enumerate}

\smallskip\noindent\textbf{MPC-based}: within this class of methods, we measure the performance of traditional optimization-based approaches using a Model Predictive Control (MPC) approach. 
\begin{enumerate}
\item \textit{MPC-Oracle}: this benchmark serves the purpose of quantifying the performance of an "oracle" controller. We provide this model with perfect foresight information of all future user requests and system dynamics within the planning horizon $K$. However, notice that this optimization model is NP-hard and thus does not scale well with increasing instance sizes. %Concretely, we optimize the desired vehicle distribution using a tri-level embedded optimization model. The upper-level model determines the desired vehicle distribution, and the lower-level models are the matching problem and the minimum rebalancing-cost problem, which characterize the impact of the optimal desired vehicle distribution on the other modules. 
\item \textit{MPC-Forecast}: we relax the assumption of perfect foresight information in MPC-Oracle and substitute it with a noisy and unbiased estimate of demand. This method serves as a realistic control-based benchmark in situations where system dynamics are unknown. As for MPC-Oracle, this formulation suffers from the same lack of scalability. 
\end{enumerate}

\begin{table*}
\centering
\begin{tabular}{l | c c c | c c c c}
    \hline
    & \multicolumn{3}{c|}{Performance} & \multicolumn{4}{c}{Dataset characteristics} \\
   
    Dataset & Expert & Behavior Policy & CQL & Reward & Spread & IQR  \\
    \hline
    NYC Brooklyn-M & 56.1 & 42.6$(\pm{2.4})$ & \textbf{53.1}($\pm{1.1}$) & \phantom{0}76\% & 0.42 & 0.24 \\
    NYC Brooklyn-H & 56.1 & 50.2($\pm{1.3}$) & 52.0($\pm{1.3}$) & \phantom{0}89\% & 0.16 & 0.13 \\
    NYC Brooklyn-G & 56.1 & 49.6($\pm{0.9}$) & 50.9($\pm{1.4}$) & \phantom{0}88\% & 0.03 & 0.03 \\
    NYC Brooklyn-E & 56.1 & 56.3($\pm{0.9}$) & 56.4($\pm{0.8}$) & 100\% & 0.12 & 0.11\\
    \hline
    Shenzhen West-M & 62.3 & 45.8$(\pm{3.6})$ & \textbf{61.6}$(\pm{0.5})$ & \phantom{0}74\% & 0.50 & 0.27 \\
    Shenzhen West-H & 62.3 & 56.1$(\pm{1.6})$ & 59.7$(\pm{0.9})$ & \phantom{0}90\% & 0.25 & 0.15 \\
    Shenzhen West-G & 62.3 & 55.9$(\pm{0.7})$ & 57.8$(\pm{1.5})$ & \phantom{0}89\% & 0.03 & 0.03\\
    Shenzhen West-E & 62.3 & 62.0($\pm{0.9}$) & 61.9($\pm{0.7}$) & 100\% & 0.13 & 0.11 \\
    \hline
    San Francisco-M & 14.8 & 11.6$(\pm{1.0})$ & 14.2$(\pm{0.4})$ & \phantom{0}78\% & 0.64 & 0.46\\
    San Francisco-H & 14.8 & 13.6$(\pm{0.4})$ &\textbf{14.3}$(\pm{0.4})$ & \phantom{0}91\% & 0.47 & 0.27 \\
    San Francisco-G & 14.8 & 13.1$(\pm{0.5})$ & 13.6$(\pm{0.5})$ & \phantom{0}88\% & 0.04 & 0.04 \\
    San Francisco-E & 14.8 & 14.7($\pm{0.3}$) & 14.8$(\pm{0.3})$ & 100\% & 0.28 & 0.24 \\
    \hline
\end{tabular}
\caption{Performance of offline agents trained on different datasets and comparison of dataset characteristics. The best-performing model not trained on expert data is highlighted in bold. Reward indicates the relative performance of the behavior policy compared to the expert policy.}
\label{tab:tab2}
\end{table*}

\subsection{Learning from Offline Data}
\label{subsec:learning_from_offline_data}
We evaluate the rebalancing algorithms on taxi trip data from the cities of NYC (Brooklyn), Shenzhen (Downtown West), and San Francisco. Please refer to the Appendix for data description and training specifics. 

%Together with the accumulated reward within the transportation network (i.e., profit), we also monitor the number of served customers and the cost of rebalancing trips as key metrics within AMoD systems. 
%We test the deterministic polices for $10$ episodes and report the mean and standard deviation in Table \ref{tab:tab1}. 
Results in Table \ref{tab:tab1} show that by learning from a static dataset collected by a sub-optimal policy, the RL agent achieves performance on par with online methods, outperforming both heuristics and optimization-based benchmarks. Specifically, AMoD control policies learned through offline RL are only $7.2\%$ (NYC), $5.5\%$ (Shenzhen), and $10.7\%$ (San Francisco) from Oracle performance, which assumes perfect information about future system dynamics. More importantly, the resulting policies are only $5.3\%$ (NYC), $1.1\%$ (Shenzhen), and $4\%$ (San Francisco) from the performance of their online counterpart, which is allowed to freely interact with the urban network during training. Results in Table \ref{tab:tab1} also show how both RL-based approaches can achieve performance that is comparable or superior to that of MPC-Forecast, thus highlighting the benefits of using learning-based approaches to control stochastic, unknown systems, a characteristic that is particularly relevant in real-world settings where accurately modeling transportation demand is in itself a challenging process. 
 
Crucially, the results highlight a drastic improvement in the sample efficiency of offline RL methods. If, on one hand, learning from online data requires approximately between $20\,000$ and $200\,000$ interactions with the transportation network, the offline agent is solely trained on a fixed dataset of $10\,000$ samples, resulting in up to $20\times$ improvement in sample efficiency. Most importantly, the samples used by the offline approach do not result from active interaction with the system but rather from historical data readily available to any service operator, thus challenging the current paradigm of simulation-based training and online interaction.

\subsection{Impact of Dataset Quality}
\label{subsec:impact_of_dataset_quality}
Within the offline RL literature, the fact that the performance of offline learning is highly dependent on the dataset characteristics is well-established \cite{qin2022neorl, schweighofer2021understanding}. 
Hence, we conduct experiments using a variety of datasets by collecting training data from policies with different degrees of optimality \cite{qin2022neorl, D4RL}. We store data generated by policies with approximately $75\%$ (Medium \textit{M}) and $90\%$ (High \textit{H}) and $100\%$ (Expert \textit{E}) of the online RL performance. Additionally, we collect a dataset via a deterministic, greedy heuristic, denoted by Greedy (\textit{G}). The heuristic selects rebalancing trips to match the projected demand distribution averaged over the next $K$ time steps. 

The results reported in Table \ref{tab:tab2} confirm the strong dependency of offline learning on the dataset characteristics. Interestingly, a higher performance of the behavior policy does not indicate a better performance of the offline agent. Specifically, the policy trained on the Medium dataset outperforms the one trained on High in both NYC and Shenzhen scenarios. Moreover, despite being generated by a deterministic heuristic with a narrow policy, the G-dataset results in control policies that outperform the policy used to generate the data it was trained on. For example, the policy obtained via CQL in the Shenzhen scenario achieves improved rebalancing cost and revenue by $11\%$ and $1\%$, respectively, compared to the greedy heuristic. The results demonstrate that all policies outperform the No Rebalancing and heuristic baselines, except for the agent trained on San Francisco-G and Shenzhen-G, suggesting that offline  agents can produce policies that can be reliably deployed.

As an additional analysis, we report the relative reward of the behavior policy, together with the spread and 0.01-/0.99-interquartile range (IQR) of the actions in the dataset. This information is used to report the coverage (i.e., diversity) of the dataset. It should be noted that determining coverage in continuous state-action spaces is still an open challenge \cite{schweighofer2021understanding}; however, we believe the IQR to be a reasonable metric across the problem settings we investigate. This information provides insights into the inferior performance of the High- and Greedy-datasets for NYC and Shenzhen. Both datasets exhibit a narrow action distribution while still being far from expert performance. Hence, the agents are implicitly constrained to a limited subset of actions by the conservative loss on the Q-function, limiting the possibilities for improvement. Low coverage in the dataset can only be compensated by the high performance of the behavior policy, as in the case of the Expert dataset. Following the same line of reasoning, the superior performance of the Medium dataset can also largely be attributed to its diversity, where the difference in dataset coverage is a result of the method used for data collection. Specifically, as the training progresses, the Dirichlet distribution becomes increasingly narrow and the agent less exploratory. %This results in the High dataset being characterized by a better-performing behavior policy but a considerably less diverse dataset. An exception is San Francisco, where the High dataset shows a diversity similar to the Medium dataset of the other cities, confirming our intuition and thus explaining the good performance of the CQL agent. 
This finding is coherent with prior work, where CQL generally performs best on datasets with high coverage \cite{schweighofer2021understanding}. More broadly, we find that a fixed policy, which hardly explores different regions of the state-action space, is likely to be sub-optimal for offline learning. In light of this, current operators should consider ways of gathering operational data while keeping in mind this quality-diversity trade-off.

\subsection{Transfer and Generalization}
\label{subsec:transfer_and_generalization}
To extensively assess the generalization capabilities of our proposed approach, we also study the extent to which policies learned through offline data are able to generalize to different cities. Specifically, the ability to transfer policies learned through offline RL to new environments without the need for additional data is particularly valuable for operators looking to expand their service to new cities. Table \ref{tab:tab4} presents the results of evaluating the zero-shot performance of a policy trained on offline data for the NYC Brooklyn dataset in comparison to (i) an agent trained online on NYC Brooklyn (i.e., SAC) and (ii) an expert policy, which is fully re-trained in the respective city (i.e., Expert). 
With zero-shot, we refer to the case in which policies are trained on data belonging to one city (e.g., New York) and later deployed to another system (e.g., Rome) \textit{without further training}. The results show that the policy learned by CQL exhibits similar or better generalization performance compared to SAC trained on online data. Most importantly, despite being trained offline on data from a different city, agents learned via offline RL achieve an interesting degree of portability to new cities compared to the Expert policy, which has been re-trained from scratch in every environment. As observed in the previous sections, these experiments confirm how the transfer capabilities of offline policies are also highly dependent on the quality-diversity trade-off. This is evident in the case of the expert dataset (characterized by the smallest coverage but the highest reward), which leads to a policy that, although achieving the best performance in NYC, is outperformed by the CQL-M agent in other cities. In other words, learning from datasets that are more diverse, even if less performant, leads to agents that are better able to generalize across a variety of scenarios. 

\begin{table}
    \centering\fontsize{9}{11}\selectfont
\begin{tabular}{lr|ccc}
      \hline
    City&Expert&SAC&CQL-M&CQL-E\\
      \hline
    Rome&3.2&\textbf{2.9}($\pm{0.2}$)&2.8($\pm{0.2}$)&2.7($\pm{0.2}$)\\
    San Fran.&14.7&14.0($\pm{0.5}$)&\textbf{14.2}($\pm{0.4}$)&13.9($\pm{0.5}$)\\
    Shz West&62.0&58.3($\pm{3.1}$)&\textbf{60.0}($\pm{1.0}$)&59.3($\pm{1.2}$)\\
    Wash. DC&13.4&13.0($\pm{0.8}$)&\textbf{13.2}($\pm{0.3}$)&12.9($\pm{0.5}$)\\
      \hline
\end{tabular}
\caption{Zero-shot performance of SAC and CQL trained on NYC-Brooklyn.}
\label{tab:tab4}
\end{table}

%\begin{figure*}
%  \centering
%  \begin{minipage}[b]{\columnwidth}
%    \centering
%    \includegraphics[width=\columnwidth]{Figures/nyc.png}
%    \caption{Training reward obtained by an online agent compared online fine-tuning of CQL-M in the NYC Brooklyn environment}
%    \label{fig:NYC_train}
%  \end{minipage}
%  \hfill
%  \begin{minipage}[b]{\columnwidth}
%    \centering
%   \includegraphics[width=\columnwidth]{Figures/shz.png}
%    \caption{Training reward obtained by an online agent compared online fine-tuning of CQL-M in the Shenzhen environment.}
%    \label{fig:Shz_train}
%  \end{minipage}
%\end{figure*}

\subsection{Economically-feasible Reinforcement Learning}
In this section, we quantify the practical benefits of offline learning within AMoD systems compared to online RL methods. Specifically, although technically feasible, online data collection with a partially-trained policy in AMoD systems is not realistic. Especially during the initial phases of training, the online agent requires extensive exploration of the environment, which will almost certainly result in erroneous and extremely high-cost rebalancing decisions. Within this context, offline RL offers a practically-viable alternative by avoiding interactions with the environment until a sufficient level of performance is achieved.
%At a high level, one of the main reasons behind the lack of real-world deployment of RL-based solutions resides in the cost of data collection. Specifically, although technically feasible, online data collection with a partially-trained policy in AMoD systems is not realistic. because of the potential risk imposed by a partially-trained agent controlling the mobility system. 

\begin{figure}
    \includegraphics[width=\columnwidth]{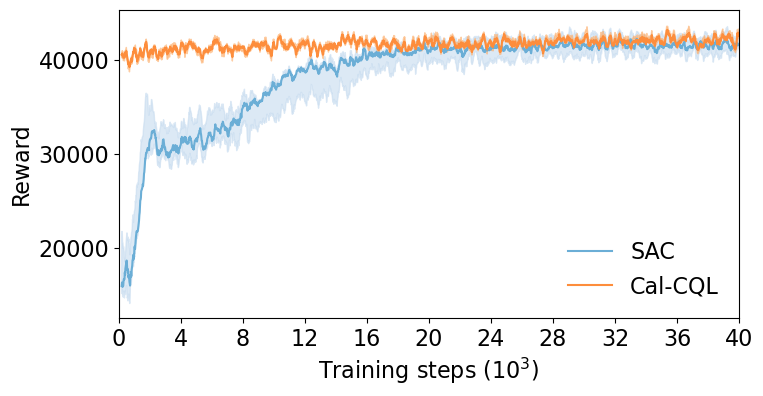}
    \caption{Training reward obtained by an online agent compared to online fine-tuning of CQL-M in the NYC Brooklyn environment}
    \label{fig:NYC_train}
\end{figure}   

%\begin{figure}
%   \includegraphics[width=\columnwidth]{Figures/shz.png}
%    \caption{Training reward obtained by an online agent compared online fine-tuning of CQL-M in the Shenzhen environment.}
%    \label{fig:Shz_train}
%\end{figure}

The training curve depicted in Figure \ref{fig:NYC_train} clearly show the extreme sub-optimality of the online learning paradigm. The offline learned agent successfully starts online fine-tuning in high-reward regions and avoids the low-reward regions that the SAC agent visits during initial exploration. Moreover, the inherent advantage of parametric policies - their tendency to become increasingly deterministic as training progresses - means that our offline pre-trained policy is inherently less stochastic, leading to safer online fine-tuning.

To further assess the potential savings offered by our approach, we compare the costs associated with training from scratch against the online fine-tuning of the offline trained agent in Table \ref{tab:savings}. We evaluate the difference in rebalancing costs and lost profit due to unmet demand using prices and costs estimated from real-world trip data.  In NYC Brooklyn, the offline agent saves $20\,780$ interactions, over $2\,600\,000\$$ in rebalancing cost, and over $160\,000\$$ of unmet demand, thus showing the significant amount of economical cost of training an agent within real-world systems. In the case of Shenzhen West, the pre-trained agent renders the entirety of online training from scratch redundant. 
Crucially, these experiments clearly quantify the benefits of offline learning and challenge the current approach of online learning by demonstrating its limitations. 

\begin{table}
     \centering\fontsize{9}{11}\selectfont
    \begin{tabular}{lccc}
          \hline
         %City & Interactions &Rebalancing Cost& Unserved demand\\
        \multicolumn{1}{c}{City}&\multicolumn{1}{c}{Interactions}&\multicolumn{1}{c}{Reb. Cost}&\multicolumn{1}{c}{Unserv. demand} \\
          \hline
         NYC Brooklyn  & 20\,780 & \phantom{0}2\,640\,100 &  \phantom{0}162\,859\\
         Shenzhen West &   200\,000 & 18\,169\,878 & 1\,720\,820\\
         San Francisco & 14\,700   & \phantom{00}174\,782 &  \phantom{0}\phantom{0}37\,899\\
          \hline
    \end{tabular}
    \caption{Cost of training an agent fully online compared to online fine-tuning of CQL-M.}
    \label{tab:savings}
\end{table}

\section{Conclusion}
\label{sec:Conclusion}
Research on the central system-level control of AMoD systems focuses on developing RL approaches via online data. However, this is hardly practical within real-world urban systems, where the cost of making wrong decisions at a fleet level is extremely high.
In this work, we propose offline RL as an appealing paradigm to approach the challenges in this space and enable centralized RL-based solutions to be applied even without simulated environments.
Specifically, we introduce an algorithm that leverages the combination of offline RL and optimization to learn how to control AMoD systems solely via offline data, thus completely removing all interactions with both the real world and complex mobility simulators. 
Our approach shows strong performance in all real-world settings we evaluate, outperforming both optimization-based and heuristic approaches and matching the performance of online RL algorithms. Crucially, our method shows sample-efficient online-fine tuning capabilities, effectively enabling service operators to avoid expensive rebalancing decisions that make online training an impractical solution within real-world settings.

In future work, we plan to investigate the combination of offline RL with ways to explicitly consider transfer and generalization in the design of neural architectures and training strategies, such as casting the problem under the lens of Offline Meta-RL \cite{NEURIPS2021_24802454}. More generally, we believe this research opens several promising directions for the application of these concepts within economically-critical, real-world mobility systems.

\bibliography{aaai24}

\appendix
\onecolumn
\section{Appendix}
First, we provide the specifics of the three-step framework employed to address the AMoD control problem. Following that, we outline the experimental details, encompassing data statistics and training parameters. Lastly, we highlight our primary finding from training the online baseline, which is the significance of the critic architecture, and offer supplementary materials with additional online fine-tuning training curves.
\subsection{The Three-step Framework}
\label{subsec:3-step-framework}
As in \cite{Gammelli1, Gammelli2}, this paper adopts a three-step decision-making framework to tackle the AMoD control problem. This framework comprises three stages: (1) solving a matching problem to derive the passenger flow, (2) determining the desired distribution of idle vehicles through the use of the learned policy $\pi_\phi(a_t|s_t)$, (3) converting this distribution to a rebalancing flow by solving a minimal rebalancing-cost problem. The advantage of this procedure is the reduction of the action space from $|\mathcal{V}|^2$ to $|\mathcal{V}|$, as the policy defines the action at each node as opposed to along each edge (i.e., each origin-destination pair).

In the first step, vehicles are assigned to customers by solving the following assignment problem to derive passenger flows $\{x^t_{i,j}\}_{i,j \in \mathcal{V}}$:
\begin{subequations}
\begin{align}
    \max_{\{x_{ij}^t\}_{i,j\in\mathcal{V}}} \quad& \sum_{i,j\in\mathcal{V}}x_{ij}^t(p_{ij}^t- c_{ij}^t) \label{eq:matching_obj}\\
    \rm{s.t.} \,\,\,\,\, \quad& 0\leq x_{ij}^t \leq d_{ij}^t, ~i,j\in\mathcal{V}, \label{eq:matching_con1}\\
    & \sum_{j \in \mathcal{V}} x_{ij}^t \leq M_{i}^t, ~i\in\mathcal{V}, \label{eq:matching_con2}
\end{align}\label{eq:matching}
\end{subequations}
where objective function \eqref{eq:matching_obj} maximizes the total profit, which is computed as the difference between revenue and the cost of the matched trips. Constraint \eqref{eq:matching_con1} ensures that the resulting passenger flow is non-negative and upper-bounded by the demand, while constraint \eqref{eq:matching_con2} guarantees that the assigned passenger flow does not exceed the number of available vehicles $M_i^t$ at station $i$ at time step $t$. Note that since the constraint matrix is totally unimodular, the resulting passenger flows are integers as long as the demand is integral. 

 Within the second step, the learned policy $\pi_\phi(a_t|s_t)$ determines the desired idle vehicle distribution $a^t_{reb} = \{a^t_{reb,i}\}_{i \in \mathcal{V}}$, where $a^t_{reb,i} \in [0,1]$ defines the percentage of currently idle vehicles to be rebalanced towards station $i$ in time step $t$, and $\sum_{i \in \mathcal{V}}a^t_{reb,i}=1$. Given the desired vehicle distribution, we denote the number of desired vehicles as $\hat{m}^t_i = \lfloor a^t_{reb,i}\sum_{i \in \mathcal{V}}m^t_i \rfloor $, where $m^t_i$ represents the actual number of idle vehicles in region $i$ at time step $t$. The use of the floor operation ensures that the desired number of vehicles is an integer and is constrained by the actual number of idle vehicles. ($\sum_{i \in \mathcal{V}}\hat{m}^t_i \leq \sum_{i \in \mathcal{V}}m^t_i$)
 
 The third step converts the desired distribution into rebalancing flows $\{y^t_{i,j}\}_{i,j \in \mathcal{E}}$ by solving a minimal rebalancing cost problem:  
\begin{subequations}
\begin{align}
    \min_{\{y_{ij}^t\}_{i\neq j\in \mathcal{V}}\in \mathbb{N}^{|\mathcal{V}|\times (|\mathcal{V}|-1)}} \quad& \sum_{i\neq j\in \mathcal{V}}c_{ij}^t y_{ij}^t \label{eq:reb_obj}\\
    \rm{s.t.}  \quad \quad \quad \,
    & \sum_{j\neq i}(y_{ji}^t - y_{ij}^t) + m_i^t \geq \hat{m}_i^t , ~i\in\mathcal{V},\label{eq:reb_con1} \\
    \quad& \sum_{j\neq i} y_{ij}^t \leq m_i^t,~i\in\mathcal{V},\label{eq:reb_con2}
\end{align}\label{eq:reb}
\end{subequations}
where objective function \eqref{eq:reb_obj} minimizes the rebalancing cost, constraint \eqref{eq:reb_con1} ensures that the resulting number of vehicles is close to the desired number of vehicles, and constraint \eqref{eq:reb_con2} limits total rebalancing flow from a region to the number of idle vehicles in that region. 

\subsection{Experimental Details}
All RL modules were implemented using PyTorch \cite{Pytorch} and the IBM \cite{IBM} CPLEX solver for the dispatching and minimal rebalancing cost problems. The offline learning experiments were conducted on a Tesla V100 GPU. 

We train each online baseline model for $10\,000$ episodes with a time horizon $T=20$, resulting in $200\,000$ training steps. As a planning horizon, we set $K=6$. For our offline RL experiments, we train each model on a fixed dataset of $10\,000$ transitions.

The results reported in the main paper show the mean and standard deviation over 10 episodes of interaction with the environment. Additionally, for all RL-based methods, the reported mean and standard deviation are for three different runs with seeds 5, 10 and 42.

\label{subsec:A1}
\paragraph{Data} The case studies are based on the taxi record data collected in San Francisco \cite{PiorkowskiSarafijanovicEtAl2009}, Washington DC \cite{WashingtonDC2019}, New York City \cite{TaxiLimousineCommission2013}, Rome \cite{BraccialeBonolaEtAl2014}, and Shenzhen \cite{ZhangZhaoEtAl2015}. As in \cite{Gammelli2}, the two larger cities, New York City and Shenzhen, are further divided into four regions such that the inter-regional demand is minimized, where we picked one region to test our approach. In each scenario, the road network is segmented into stations by clustering junctions such that the travel time within each station is upper-bounded by a given error tolerance (e.g., a time step). The trip record data are converted to demand, travel times, and trip prices between stations. The demand (Poisson rate of customer arrival) is aggregated from the trip record data every 15 minutes. Descriptive statistics of the scenarios used are presented in Table \ref{tab:statistics}.

\begin{table}[h]
\centering
  \caption{Scenario statistics}
  \label{tab:statistics}
  \begin{tabular}{ccccccc}
    \hline
    City &Date and time&No. Nodes & \thead{Max. Trip Time\\(Min)}& \thead{Avg. Trip Time\\(Min)} &\thead{Avg. Demand\\(req./hr)}&No. Vehicles\\
    \hline
    NYC Brooklyn&2013-03-08 19:00-20:00&14&68&18&3162&1500\\
    Shenzhen West&2013-10-22 8:00-9:00&17&66&15&4637&1777\\
    San Francisco&2008-06-06 8:00-9:00&10&55&11&1380&374\\
    Rome&2014-02-04 8:00-9:00&13&69&18&177&79 \\
    Washington DC& 2019-03-12 19:00-20:00&18&68&14&1000&1097\\
  \hline
\end{tabular}
\end{table}

\paragraph{SAC specifications} The hyperparameters used to train our online baseline can be found in Table \ref{tab:hyperSAC}. We deploy the double Q-trick and target Q-networks. 
\begin{table}[H]
\centering
  \begin{tabular}{l|c}
    \hline
    Parameter &Value\\
    \hline
    Optimizer & Adam\\
    Learning rate & $1*{10^{-3}}$\\
    Discount ($\gamma$) & 0.97\\
    Replay buffer size & 200000\\
    Batch size& 100\\
    Entropy coefficient& 0.3\\
    Target smoothing coefficient ($\tau$)&0.005\\
    Target update interval&1\\
    Gradient step/env.interaction &1\\
    Reward scaling& 0.01 (for SF 0.1)\\
  \hline
\end{tabular}
  \caption{Hyperparameters SAC}
  \label{tab:hyperSAC}
\end{table}

\paragraph{CQL specifications} For our offline experiments, we train the $CQL(\mathcal{H})$ version of CQL and estimate the log-sum-exp using importance sampling, where we sample 10 actions from a Dirichlet distribution with concentration parameter $\textit{c} = [1,1,..,1]$ and the current policy $\pi$. We use a policy learning rate of $1*10^{-4}$ and a critic learning rate of $3*{10^{-4}}$. Our experiments revealed that the stability of the training on certain datasets can be improved by reducing the reward scaling, as indicated in Table \ref{tab:hyperCQL}. 

The regularizer weight $\eta$ can also be chosen via dual-gradient-descent by introducing a threshold $\tau$ \cite{Kumar}. 
\begin{equation}
\min_{Q} \max_{\eta \geq 0} \eta \Bigl( \mathbb{E}_{s\sim D} [log \sum_{a} exp(Q(s,a)) - \mathbb{E}_{a\sim \hat{\pi}_{\beta}(a|s)}[Q(s,a)]] - \tau \Bigr)+TD Error\nonumber,
\end{equation}
where a larger value of $\eta$ is assigned if the difference between Q-values exceeds $\tau$ to more aggressively penalize Q-values. Default values were $\eta = 1$ and $\tau = -1$ ($\tau = -1$ denotes the usage of a fixed $\eta$). Further settings were tried to improve the stability of losses and q-values if needed, specifically  $\eta =\in \{1,2, 5\}$ and $\tau \in \{-1, 5, 10\}$. The results are reported in Table \ref{tab:hyperCQL}. The remaining hyperparameters are kept identical to the online SAC version in Table \ref{tab:hyperSAC}. 
\begin{table}
\centering
  \caption{Hyperparameters CQL}
  \label{tab:hyperCQL}
  \begin{tabular}{l|cccc}
    \hline
    Task name &$\eta$&$\tau$&reward scaling\\
    \hline
    NYC Brooklyn-Medium &1&10&0.001 \\
    NYC Brooklyn-High &1&10&0.01\\
    NYC Brooklyn-Heuristic &1&-1&0.01\\
    NYC Brooklyn-Expert &5&-1&0.01\\
    \hline
    Shenzhen West-Medium &1&-1&0.01\\
    Shenzhen West-High &1&10&0.001\\
    Shenzhen West-Heuristic &1&5&0.001\\
    Shenzhen West-Expert &5&-1&0.001\\
    \hline
    San Francisco-Medium &1&10&0.1\\
    San Francisco-High &1&10&0.1\\
    San Francisco-Heuristic &2&-1&0.1\\
    San Francisco-Expert &5&-1&0.01\\
  \hline
\end{tabular}
\end{table}

\paragraph{Cal-CQL specifications} For the online fine-tuning, we use the same hyperparameters as for the SAC online training with the exception that during training, we sample $25\%$ of the batch from the offline dataset the Cal-CQL agent was trained on and $75\%$ from the online replay buffer. We train the Cal-CQL agents offline with $\eta=1$ and $\tau=-1$.
\subsection{Online Training}
\label{subsec:A2} 
While training the online baselines, we made the finding that the critic architecture is essential for achieving satisfactory results. We tested six critic architectures and determined a group of architecture designs that is more favorable to work with a graph-based state and action. We tested different ways to represent the action as an input to the Q-function (i.e., (i) as an additional input feature for the GCN, (ii) via concatenation, or (iii) via pointwise multiplication after GCN): 
\begin{enumerate}
\item \textit{Critic1}: first the GCN encodes the state information, and its output is point-wise multiplied with the action. A readout layer (global sum) is applied and three fully connected layers are applied on the graph level.
\item \textit{Critic2}: first the GCN encodes the state information. Then, a readout layer (global sum) is applied and its output is concatenated with the action. Three fully connected layers are applied.
\item \textit{Critic3}: the action is added as an additional node feature for the GCN and its output is aggregated using a readout layer (global sum). The fully connected layers are applied.
\item \textit{Critic4}: first the GCN encodes the state information and its output is concatenated with the action. Two fully connected layers are applied. The readout layer is applied before the final fully connected layer. 
\item \textit{Critic1 (v2)}: first the GCN encodes the state information, and its output is point-wise multiplied with the action. The output is processed by two fully connected layers. The readout layer is applied before the final fully connected layer. 
\item \textit{Critic3 (v2)}: the action is added as an additional node feature for the GCN. The output is processed by two fully connected layers. The readout layer is applied before the final fully connected layer.
\end{enumerate}

Our findings indicate that the most critical aspect is the joint processing of both action and node features at the node level prior to their aggregation at the graph level. This can be achieved by either incorporating the action information into the input node features of the GNN or applying two fully connected layers before aggregating the information with the readout layer.

As demonstrated in Figure \ref{fig:critics}, it is evident that the two architectures that do not process the states and actions jointly at the node level (\textit{Critic1} and \textit{Critic2}) fail to produce a meaningful estimate of the expected reward. Consequently, the actor is unable to learn an effective policy. Moreover, we found that the architecture that inputs the action information to the GCN but performs the readout directly thereafter (\textit{Critic3}) can be further improved by applying the MLP at the node level, resulting in \textit{Critic3}(v2).

%\begin{figure}[!h]
%    \centering
%    \includegraphics[width=0.8\textwidth]{Figures/critics.png}
%    \caption{Impact of different critic architectures}
%    \label{fig:critics}
%\end{figure}

\subsection{Additional Material}
\begin{figure*}
  \centering
  \begin{minipage}[b]{0.49\columnwidth}
    \centering
    \includegraphics[width=\columnwidth]{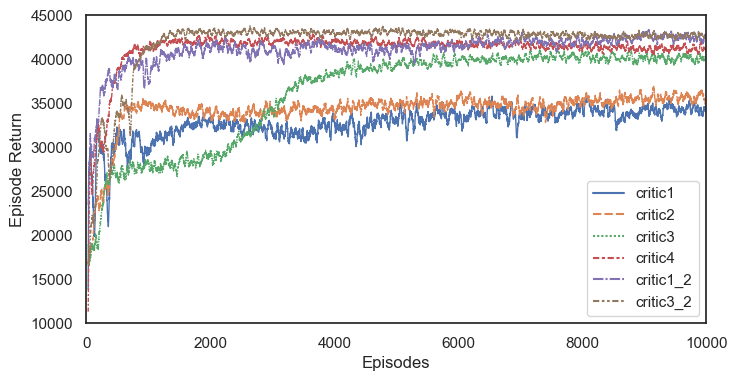}
    \caption{Impact of different critic architectures on the training reward while online training in the NYC Brooklyn environment}
    \label{fig:critics}
  \end{minipage}
  \hfill
  \begin{minipage}[b]{0.49\columnwidth}
    \centering
   \includegraphics[width=\columnwidth]{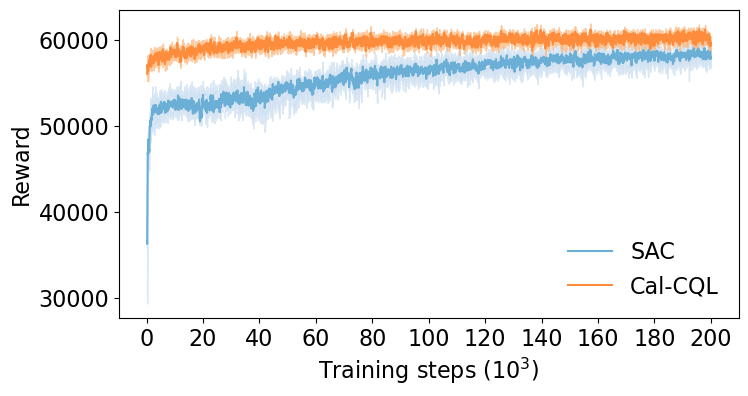}
    \caption{Training reward obtained by an online agent compared to online fine-tuning of CQL-M in the Shenzhen environment.}
    \label{fig:Cal_SHZ}
  \end{minipage}
\end{figure*}

\begin{figure*}
  \centering
  \begin{minipage}[b]{0.49\columnwidth}
    \centering
    \includegraphics[width=\columnwidth]{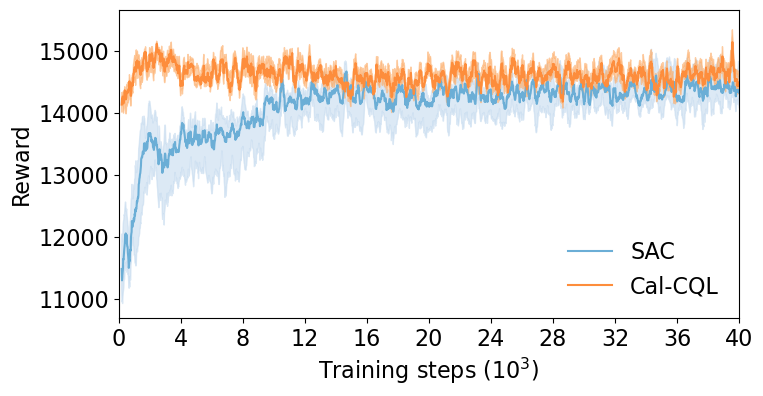}
    \caption{Training reward obtained by an online agent compared to online fine-tuning of CQL-M in the San Francisco environment.}
    \label{fig:Cal_SF}
  \end{minipage}
  \hfill
  \begin{minipage}[b]{0.49\columnwidth}
    \centering
   \includegraphics[width=\columnwidth]{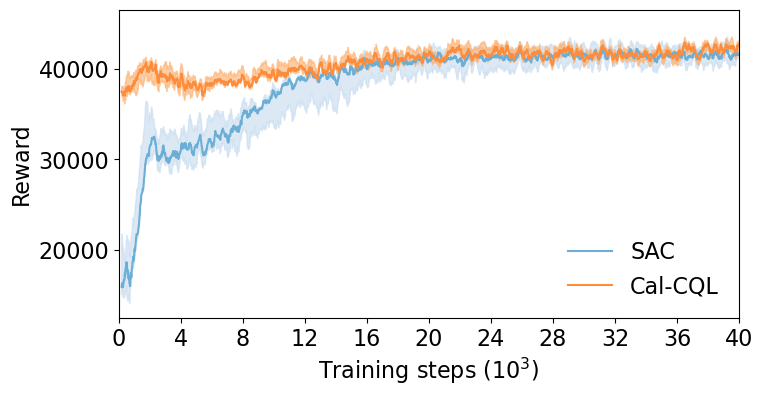}
    \caption{Training reward obtained by an online agent compared to online fine-tuning of CQL-G in the NYC environment.}
    \label{fig:Cal_NYC_G}
  \end{minipage}
\end{figure*}

In this section, we supply additional material that complements Section Transfer and Generalization and Section Economically-feasible RL. Figures \ref{fig:Cal_SHZ} and  \ref{fig:Cal_SF} depict the training curves of online fine-tuning for the Cal-CQL agent and the online training from Scratch for the SAC agent for Shenzhen downtown west and San Francisco. Additionally, we provide the online fine-tuning training curves of the Cal-CQL agent trained on the Greedy dataset for NYC Brooklyn in Figure \ref{fig:Cal_NYC_G}. 

Table \ref{tab:addbaselines} provides additional baselines for the cities of Rome and Washington DC that were used to zero-shot test the offline trained policies in Table \ref{tab:tab4}. 

%\begin{figure*}[!h]
%  \centering
%  \begin{minipage}[b]{0.49\columnwidth}
%    \centering
%    \includegraphics[width=\columnwidth]{Figures/nyc_g2.png}
%    \caption{Training reward obtained by an online agent compared to online fine-tuning of CQL-G in the NYC environment.}
%    \label{fig:Cal_NYC_G}
%  \end{minipage}
%\end{figure*}

\begin{table}[!h]
\centering
  \caption{Average reward (profit, thousands of dollars)}
  \label{tab:addbaselines}
  \begin{tabular}{l|ccc|cc|c}
    \hline
    City&Random&No Rebalancing &ED&SAC\\
    \hline
    Rome &2.1($\pm{0.21}$)&2.3($\pm{0.2}$)&2.9($\pm{0.2}$)&3.2($\pm{0.2}$)\\
    Washington DC &9.0($\pm{0.9}$)&11.4($\pm{0.3}$)&13.0($\pm{0.2}$)&13.4($\pm{0.2}$)\\
  \hline
\end{tabular}
\end{table}

\end{document}